\documentclass[onecolumn,prl]{revtex4}
\usepackage{bm}
\usepackage{epsfig}
\usepackage{amsmath}
\begin{document}
\title{Diffusion-controlled intrachain reactions in polymer: An analytical model}
\author{Aniruddha Chakraborty \\
School of Basic Sciences, Indian Institute of Technology Mandi,\\
Mandi, Himachal Pradesh, 175001, India}
\date{\today }
\begin{abstract}
\noindent  A theory of diffusion-controlled intramo1ecular reactions of polymer chain in dilute solution is formulated. Our model is based on the widely used
diffusion-reaction formalism of Wilemski and Fixman (J. Chem. Phys. 60, 866 (1974). Our model is more general than the model proposed by Wilemski and Fixman, in the sense that our model considers the motion  of both close and open chain polymer explicitely. It is quite unlikely that the motion of close chain polymer do not play any role in the loop formation dynamics, but unfortunately this fact was not considered in any of the earlier studies.
\end{abstract}
\maketitle
Understanding the dynamics of loop formation in long chain molecules has been an interesting both, to experimentalists 
\cite{Winnik,Haung,Lapidus, Hudgins} and theoreticians \cite{Fixman,Doi,Szabo,Zwanzig,Portman,Sokolov,Thiru}. Advances in single molecule spectroscopic techniques have made it possible to monitor the kinetics of loop formation at the single molecule level \cite{Widom,Hyeon}. Loop formation is believed to be an important step in protein folding \cite{Eaton} and RNA folding \cite{Klimov}. Loop formation in polymers is actually a very complex problem and exact analytical solution for dynamics of loop formation is impossible. All the
theories of loop formation dynamics are approximate \cite{Fixman,Szabo}. The dynamics of a single polymer chain having reactive
end-groups may be modeled by the following Smoluchowski equation.
\begin{eqnarray}
\frac{\partial P_o({\bf \{R\}},t)}{\partial t} = {\cal L}_o P_o({\bf \{R\}},t) - k_0 S({\bf \{R\}}) P_c({\bf \{R\}},t)  \\ \nonumber
\frac{\partial P_c({\bf \{R\}},t)}{\partial t} ={\cal L}_c P_c({\bf \{R\}},t) + k_0 S({\bf \{R\}}) P_o({\bf \{R\}},t). \nonumber
\end{eqnarray}
Where $P_x({\bf \{R\}},t)$ is the distribution function of the open or close chain that has the conformation
${\bf R}$ = $R_1$, $R_2$,......$R_n$ at time $t$ where $R_i$ denotes the position of the i-th monomer in a
chain of $n$ monomers. $S({\bf R})$ is the sink function which actually models the reaction
between the ends of the polymer and is a function of end to end vector only. $k_0$ is the
rate constants for bond formation and  ${\cal L}_o$ is a differential
operator, defined as
\begin{equation}
{\cal L}_x= D_x \sum_{1=1}^{n} \frac{\partial}{\partial {\bf R}_i}.\left[\frac{\partial}{\partial {\bf R}_i}+\frac{\partial U_x({\bf R}_i)}{\partial {\bf R}_i}\right]P_x({\bf \{R\}},t),
\end{equation}
where $D_x$ is the diffusion coefficient of open or close chain and $U_x$ is the potential energy of open or close chain polymer. 
Our model is more general than the model proposed by Wilemski and Fixman, \cite{Fixman}, in the sense that our model considers the motion of both close and open chain polymer explicitely. It is quite unlikely that the motion of close chain polymer do not play any role in the loop formation dynamics, but unfortunately this fact was not considered in the earlier studies \cite{Fixman}.
In the follwoing we provide a general procedure for finding the exact analytical expression of $G({\bf R},t|{\bf R}_0,0)$. The Laplace transform ${\cal P}_x(\{{\bf R}\},s)=\int_{0}^{\infty} P_x(\{{\bf R}\},t)e^{-st} dt$ obeys
\begin{eqnarray}
[s-{\cal L}_o] {\cal P}_o(\{{\bf R}\},s)+k_0 S(\{{\bf R}\}) {\cal P}_c (\{{\bf R}\},s) = P_o({\bf R}_0) \\ \nonumber
[s-{\cal L}_c] {\cal P}_c(\{{\bf R}\},s)-k_0 S(\{{\bf R}\}) {\cal P}_o (\{{\bf R}\},s) = 0, \nonumber
\end{eqnarray}
where $P_o({\bf R}_0)=P_o(\{{\bf R}\},0)$  and $P_c(\{{\bf R}\},0)=0$.
\begin{equation}
 \left(
\begin{array}{c}
{\cal P}_o (\{{\bf R}\},s) \\
{\cal P}_c (\{{\bf R}\},s)
\end{array}
\right) = \left(
\begin{array}{cc}
s-{\cal L}_o & k_0 S(\{{\bf R}\}) \\
 - k_0 S(\{{\bf R}\}) & s-{\cal L}_c
\end{array}
\right)^{-1}
\left(
\begin{array}{c}
P_o({\bf R}_0) \\
0
\end{array}
\right)  ,
\end{equation}
Using the partition technique \cite{Lowdin}, solution of this equation can be expressed as 
\begin{equation}
{\cal P}_o(\{{\bf R}\},s)=\int d{\bf R}_0 G({\bf R},s|{\bf R}_0)P_o({\bf R}_0),
\end{equation}
where $G({\bf R},s|{\bf R}_0)$ is the Green's function defined by
\begin{equation}
G({\bf R},s|{\bf R}_0)=\left < {\bf R} \left|[s-{\cal L}_o +{k_0}^2 S[s-{\cal L}_c]^{-1}S]^{-1}\right| {\bf R}_0 \right>
\end{equation}
The above equation is true for any general $S$. This expressions simplify considerably if S is a radial Dirac Delta function located at $a$, {\it i.e.} $\delta (R-a)$. In operator notation $S$ may be written as $S= \left| a \left> \right < a \right |$. Then
\begin{equation}
G({\bf R},s|{\bf R}_0) = \left < {\bf R} \left| [s-{\cal L}_o + {k_0}^2 G^{0}_c(a,s;a) S ]^{-1} \right| {\bf R}_0 \right>,
\end{equation}
where
\begin{equation}
G^{0}_c({\bf R},s|{\bf R}_0)=\left < {\bf R} \left|[s-{\cal L}_c]^{-1}\right| {\bf R}_0 \right>
\end{equation}
and corresponds to the change in conformation of the close chain polymer starting from '${\bf R}_0$' can be found at '${\bf R}$' in the absence of any loop opening reaction.
Now we use the operator identity
\begin{equation}
[s-{\cal L}_o - {k_0}^2 G^{0}_c(a,s;a) S ]^{-1}=[s-{\cal L}_o]^{-1}-[s-{\cal L}_o]^{-1}{k_0}^2 G^{0}_c(a,s;a) S [s-{\cal L}_c - {k_0}^2 G^{0}_c(a,s;a) S]^{-1}
\end{equation}
Inserting the resolution of identity $I=\int_{-\infty}^{\infty} db \left|b \left> \right < b \right|$ in the second term of the above equation and integrating, we arrive at an equation which is similar to Lippman-Schwinger equation.
\begin{equation}
G({\bf R},s|{\bf R}_0)=G^0_o({\bf R},s|{\bf R}_0) - {k_0}^2 G^0_o({\bf R},s;a)G^0_c(a,s;a)G(a,s;{\bf R}_0),
\end{equation}
where 
\begin{equation}
G^{0}_o({\bf R},s|{\bf R}_0)=\left < {\bf R} \left|[s-{\cal L}_o]^{-1}\right| {\bf R}_0 \right>
\end{equation}
and corresponds to the change in conformation of the open chain polymer starting from '${\bf R}_0$' can be found at '${\bf R}$' in the absence of any loop forming reaction. We now put replace ${\bf R}$ by $a$ in Eq.(10) and solve for $G(a,s;{\bf R}_0)$ to get
\begin{equation}
G(a,s;{\bf R}_0)=\frac{G^0_o(a,s;{\bf R}_0)}{1+{k_0}^2 G^0_c(a,s;a)G^0_o(a,s;a)}.
\end{equation}
This when substitued back into Eq. (13) gives
\begin{equation}
G({\bf R},s|{\bf R}_0)=G^0_o({\bf R},s;{\bf R}_0) - \frac{{k_0}^2 G^0_o({\bf R},s;a)G^0_c(a,s;a)G^0_o(a,s;{\bf R}_0)}{1+{k_0}^2 G^0_o(a,s;a)G^0_c(a,s;a)}.
\end{equation}
Using this Green's function in Eq. (5) one can caluclate ${\cal P}_o(\{{\bf R}\},s)$ explicitely. Here we are interested to know the survival probability of the open chain polymer $ P_o(t) = \int d{\bf R} P_o({\bf R},t)$. It is possible to evaluate Laplace Transform  ${\cal P}_o(s)$ of $P_o(t)$ directly. ${\cal P}_o (s)$ is defined in terms of ${\cal P}_o({\bf R},s)$ by the following equation,
\begin{equation}
{\cal P}_o(s)=\left(1-\left[1+k_0^2 G^0_o(a,s;a)G^0_c(a,s;a)\right]^{-1}k_0^2 G^0_c(a,s;a)\int^{\infty}_{-\infty}d{\bf R}_0 G^0_o(a,s;{\bf R}_0)Po({\bf R}_0)\right)/(s).
\end{equation}
From the above equation we see that ${\cal P}_o(s)$ depends on $G^0_c(x_c,s;x_c)$ which is different from the models of all earlier studies \cite{Fixman,Doi,Szabo,Zwanzig,Portman,Sokolov,Thiru}. The average and long time rate constants can be found from ${\cal P}_o(s)$ \cite{Kls2}. Thus, $k^{-1}_{1}={\cal P}_o(0)$ and $k_{L}= - ($ pole of $\left[1+k_0^2 G^0_o(x,s;a)G^0_c(a,s;a)(s)\right]^{-1})$, closest to the origin, on the negative $s$ - axis, and is independent of the initial distribution but depends on $G^0_c(a,s;a)$. The expression that we have obtained for ${\cal P}_o(s)$, $k_I$ and $k_L$ are quite general and are valid for any type of polymers or loops. The same procudure will work for the opposite reaction {\it i.e.} loop opening reaction and also work for reversible case.

\end{document}